\documentclass[conference]{IEEEtran}

\usepackage[utf8]{inputenc}
\usepackage[style=ieee,isbn=false,url=false,eprint=false,maxcitenames=1,mincitenames=1]{biblatex}

\usepackage{siunitx}
\usepackage[super]{nth}
\usepackage{amsmath,amssymb,amsfonts, mathtools}
\usepackage{hyperref} 

\usepackage{algorithmic}
\usepackage{graphicx}
\usepackage{subcaption}
\usepackage{stfloats}
\graphicspath{Figures/}
\usepackage{textcomp}
\usepackage{soul}

\usepackage{enumitem}
\usepackage[ruled]{algorithm2e}
\def\BibTeX{{\rm B\kern-.05em{\sc i\kern-.025em b}\kern-.08em
    T\kern-.1667em\lower.7ex\hbox{E}\kern-.125emX}}
\usepackage{tikz}

\newcommand\submittedtext{%
  \footnotesize This work has been submitted to the IEEE for possible publication. Copyright may be transferred without notice, after which this version may no longer be accessible.}

\newcommand\submittednotice{%
\begin{tikzpicture}[remember picture,overlay]
\node[anchor=south,yshift=10pt] at (current page.south) {\fbox{\parbox{\dimexpr0.65\textwidth-\fboxsep-\fboxrule\relax}{\submittedtext}}};
\end{tikzpicture}%
}

\renewcommand\fbox{\fcolorbox{red}{white}}
\begin{document}

\title{Self-Supervised Representation Learning with Augmentations of Continuous Training Data Improves the Feel and Performance of Myoelectric Control}

\author{Shriram Tallam Puranam Raghu$^{1}$, Dawn MacIsaac$^{1}$, and Erik Scheme$^{1}$,\IEEEmembership{~Senior~Member,~IEEE}
\thanks{*This work was supported in part by NSERC Grant 2020-04776, Canada}
\thanks{Shriram Tallam Puranam Raghu, Dawn MacIsaac, and Erik Scheme are with the Department of Electrical and Computer Engineering and the Institute of Biomedical Engineering, University of New Brunswick, Fredericton, NB, E3B 5A3, Canada. 
        {Email: \tt\footnotesize stallam@unb.ca, dmac@unb.ca, escheme@unb.ca}}%
}

\maketitle
\submittednotice

\begin{abstract}

Pattern recognition-based myoelectric control is traditionally trained with static or ramp contractions, but this fails to capture the dynamic nature of real-world movements. This study investigated the benefits of training classifiers with continuous dynamic data, encompassing transitions between various movement classes. We employed both conventional (LDA) and deep learning (LSTM) classifiers, comparing their performance when trained with ramp data, continuous dynamic data, and continuous dynamic data augmented with a self-supervised learning technique (VICReg). An online Fitts' Law test with $20$ participants evaluated the usability and effectiveness of each classifier. Results demonstrate that temporal models, particularly LSTMs trained with continuous dynamic data, significantly outperformed traditional approaches. Furthermore, VICReg pre-training led to additional improvements in online performance and user experience. Qualitative feedback highlighted the importance of smooth, jitter-free control and consistent performance across movement classes. These findings underscore the potential of continuous dynamic data and self-supervised learning for advancing sEMG-PR-based myoelectric control, paving the way for more intuitive and user-friendly prosthetic devices.

\end{abstract}
Keywords - Surface electromyography pattern recognition, myoelectric control, steady-state, transitions, temporal, rejection, VICReg, self-supervised learning, representation learning, ISO 9241-9.

\pagestyle{plain} 

\section{Introduction}
\hfill \hfill

Despite emerging clinical adoption, translating robust surface electromyography pattern recognition (sEMG-PR) based myoelectric control from laboratory to real-world settings remains a formidable challenge. The inherent variability of sEMG signals, coupled with various confounding factors, contributes to this difficulty \cite{campbell_current_2020, stuttaford_reducing_2024, wang_unravelling_2024}. To address these challenges, researchers have increasingly turned to deep learning (DL) methods \cite{jabbari_emg-based_2020, zhai_self-recalibrating_2017, li_gesture_2021}. However, the success of DL models heavily depends on the quality and representativeness of the training data. 

Recognizing this limitation, researchers have begun exploring ways to enrich the training process by incorporating more diverse and representative data \cite{williams_multifaceted_2024, mukhopadhyay_experimental_2020}. However, one crucial aspect of real-world movement - its dynamic and continuous nature - often remains underrepresented in training datasets. Current sEMG-PR classifiers are typically trained using static data (where the data is recorded while the user is holding a contraction in steady state) or ramp data (where the data is recorded while the user is gradually transitioning from rest to an active class of contraction \cite{scheme_training_2013}). The data seen during goal-oriented use, however, do not exhibit such well-behaved patterns, and often include transitions between classes. This discrepancy can lead to performance degradation and transition errors during online use \cite{robertson_effects_2019}. Considering the ubiquity of dynamic movements in real-world applications, incorporating this variability into the training process of sEMG-PR models is essential.

In our previous works \cite{raghu_enabling_2024, raghu_self-supervised_2024}, we conducted an offline analysis to assess the benefits of training with continuous dynamic data containing transitions. Our results showed that training with continuous dynamic data improved the offline performance of both temporal deep recurrent models (Gated Recurrent Units (GRUs) and Long short-term memory networks (LSTMs)) and a non-temporal Linear Discriminant Analysis (LDA) classifier. Intriguingly, the results also suggested that the deep temporal models performed slightly better than the LDA when training with continuous dynamic data, but performed worse than LDA when trained with ramp data, revealing an interaction between training data characteristics and model type. Furthermore, our results demonstrated that the recurrent models, when trained with a Variance-Invariance-Covariance Regularization (VICReg) loss (a Self-Supervised Learning (SSL) method) \cite{bardes_vicreg:_2022}, outperformed all other classifiers, highlighting the potential of SSL in this context.

Although these initial offline results were promising, improvements in offline performance do not always translate to improved online performance \cite{ortiz-catalan_offline_2015}. Therefore, in this work, we conducted an online usability study using a Fitts law environment to assess the benefits of training with continuous dynamic data. The two main contributions of this work are thus: 1) demonstrating that training classifiers with continuous dynamic data, particularly the temporal LSTM, improves online user-in-the-loop performance in a virtual target acquisition test, and 2) a VICReg self-supervised learning approach with  data augmentations outperforms other methods, as measured by both Fitts law usability metrics and through qualitative feedback.

\section{Methods}

Twenty able-bodied participants ($N=20$, age: $26.9 \pm 7.9$, 12 Female, 8 Male) were recruited for this study mainly from a student population. The study was approved by the University of New Brunswick’s Research Ethics Board (REB \#2022-192). All participants provided written informed consent before participating, in accordance with the Declaration of Helsinki. Participants also completed a short survey to report their prior experience with myoelectric control systems. The distribution of self-reported experience levels was as follows: 3 reported no prior experience, 9 participants reported low experience, 4 reported medium experience, and 4 reported high experience.

Participants then completed a two-phase experiment designed to evaluate the efficacy of the different classifiers in a Fitts' Law-style target acquisition test. The first phase involved the collection of data to train and validate the classification algorithms. This phase consisted of ramp and continuous dynamic transition trials between various gestures, requiring participants to exert various muscle contractions based on visual prompts. The second phase assessed user performance in the Fitts' Law test, where participants interacted with a virtual environment controlled by the different classifiers. The details of the experiment are explained below. The software stack used for the study was as follows:
\begin{itemize}
    \item \textbf{Processing}: Numpy (v1.26) \cite{harris_array_2020}, SciPy (v1.12)\cite{virtanen_scipy_2020}, 
    \item \textbf{Pattern Recognition}: Scikit-learn (v1.4) \cite{pedregosa_scikit-learn:_2011} , Keras \cite{chollet2015keras} / TensorFlow (v2.15) \cite{tensorflow2015-whitepaper}, 
    \item \textbf{Statistical Analysis and Visualization}: Pandas (v2.2) \cite{mckinney_data_2010}, Matplotlib (v3.8) \cite{hunter_matplotlib:_2007} / Seaborn (v0.13) \cite{waskom_seaborn:_2021}, Pingouin (v0.5) \cite{vallat_pingouin:_2018}.
\end{itemize}

\subsection{Data Acquisition System}

A Delsys Trigno\textregistered\ system was used to record the sEMG signals from the forearm muscles. Six bipolar electrodes were placed equidistantly around the circumference of the forearm, one-third of the distance distally from the elbow, in a clockwise fashion starting above the middle of the flexor carpi radialis muscle. The device sampled the signals at \SI{2}{\kilo\hertz} and digitized them using a 16-bit Analog-To-Digital converter. The signals were bandpass filtered \qtyrange{20}{450}{\hertz} using a \nth{4} order zero-phase filter, and a \SI{60}{\hertz} notch filter to remove any potential noise \cite{simao_review_2019, samuel_intelligent_2019}.

\subsection{Training Data Protocol}
Data collection was facilitated by a custom Python-based GUI environment, with the protocol for recording ramp and continuous data similar to the one used in \cite{raghu_decision-change_2023}. Participants first completed five ramp trials, each consisting of contractions for seven classes: Wrist Flexion (WF), Wrist Extension (WE), Wrist Pronation (WP), Wrist Supination (WS), Hand Close (HC), Hand Open (HO), and No Movement (NM). All trials were paced via on-screen visual prompts. During each trial, participants were instructed to begin from a neutral position and ramp up to a moderate contraction intensity for each class (except NM) over a \SI{3}{\second} interval. For the NM class, participants relaxed for the entire \SI{3}{\second}. After each interval, participants released the contraction and returned to a neutral position for \SI{3}{\second} during which they were shown a preview of the upcoming contraction class. Data were obtained only from the \SI{3}{\second} ramp-up phase; the relaxation periods were not recorded. The total ramp data amounted to 5 trials x 7 prompts / trial x \SI{3}{\second} / prompts = \SI{105}{\second} per participant.

Next, participants completed six continuous dynamic trials. Each trial included transitions between all class pairs (e.g., WF to WE) resulting in a total of $7 \times 6 = 42$ transitions, and consequently, 43 steady-state regions. Unlike the ramp trials, prompts were presented in a randomized order, and participants received no rest periods or previews between prompts. They were instructed to transition to the prompted class quickly and smoothly, and maintain the contraction in steady state until the next prompt appeared on the screen.  The prompts lasted \SI{3}{\second} each. The continuous dynamic data amounted to 6 trials x 43 prompts / trial x \SI{3}{\second} / prompts = \SI{774}{\second} per participant.

These data will be made publicly available through LibEMG\cite{eddy_libemg:_2023}.

\subsection{Pattern Recognition Pipeline}

The PR pipeline used in this study mirrors that of our previous offline study \cite{raghu_self-supervised_2024}, and is briefly described below.

\subsubsection{Feature Extraction}
First, EMG data from each trial were segmented into overlapping windows or frames. Each frame had a duration of \SI{162}{\milli\second} and an increment of \SI{13.5}{\milli\second}. This increment value was chosen to match the packet rate of the Delsys Trigno system. A set of features known as Low-Sampling Frequency 4 (LSF4) features, proposed in \cite{phinyomark_feature_2018}, were then extracted from each frame, which was used for classification. Additionally, the Mean Absolute Value (MAV) was extracted from each frame for proportional control. 

\subsubsection{Labeling}
Labels for the ramp data windows were directly assigned based on the corresponding prompts presented during data collection. However, a low-amplitude threshold was applied to relabel windows with minimal EMG activity as the NM class, as described in \cite{scheme_training_2013}. Unlike ramp data, labeling continuous dynamic data frames presented a challenge due to the lack of precise information about user response timings. In prior studies \cite{raghu_decision-change_2023}, external devices like Leap Motion Controllers have been used to capture user responses. However, to avoid relying on such devices and to streamline the labeling process, we adopted an approach based on the well-established concept of visual choice reaction time (CRT) \cite{woods_age-related_2015}. 

CRT refers to the time it takes for a user to react to a stimulus and select a response. To incorporate this concept, we used the average response times, i.e., the time to initiate a movement in response to a prompt, observed in our previous study \cite{raghu_decision-change_2023}. Our data indicated a median response time of approximately $\SI{464}{\milli\second}$, which is consistent with the findings reported in \cite{woods_age-related_2015}. Therefore, we assumed this constant delay between the visual prompt presentation and the user's initiation of the corresponding muscle contraction. Correspondingly, the labels for the continuous dynamic data windows were assigned based on the presented prompts while accounting for the user delay. This approach offers a balance between accuracy and practicality, avoiding the need for time-consuming manual annotation while acknowledging inherent user response delay.

\subsubsection{Classification}
This subsection outlines the classification models utilized in this study. Building upon our prior studies on the contrasting behaviors of temporal and non-temporal models \cite{raghu_self-supervised_2024}, we evaluated five distinct classifiers

\begin{enumerate}
    \item \textbf{LDA-R}: LDA trained with ramp data is commonly used as a baseline, and was implemented as delineated in \cite{scheme_confidence-based_2013}.
    \item \textbf{LSTM-R}: An LSTM classifier trained with ramp data using cross-entropy (XEnt) loss.
    \item \textbf{LDA-D}: An LDA classifier trained with the CRT-labeled continuous dynamic data.
    \item \textbf{LSTM-D}: An LSTM classifier trained with the CRT-labeled  continuous dynamic data using XEnt loss.
    \item \textbf{LSTM-V}: An LSTM classifier pre-trained without labels using VICReg loss.
\end{enumerate}

\begin{figure*}[htbp]
    \centering
        \includegraphics[width=0.75\linewidth]{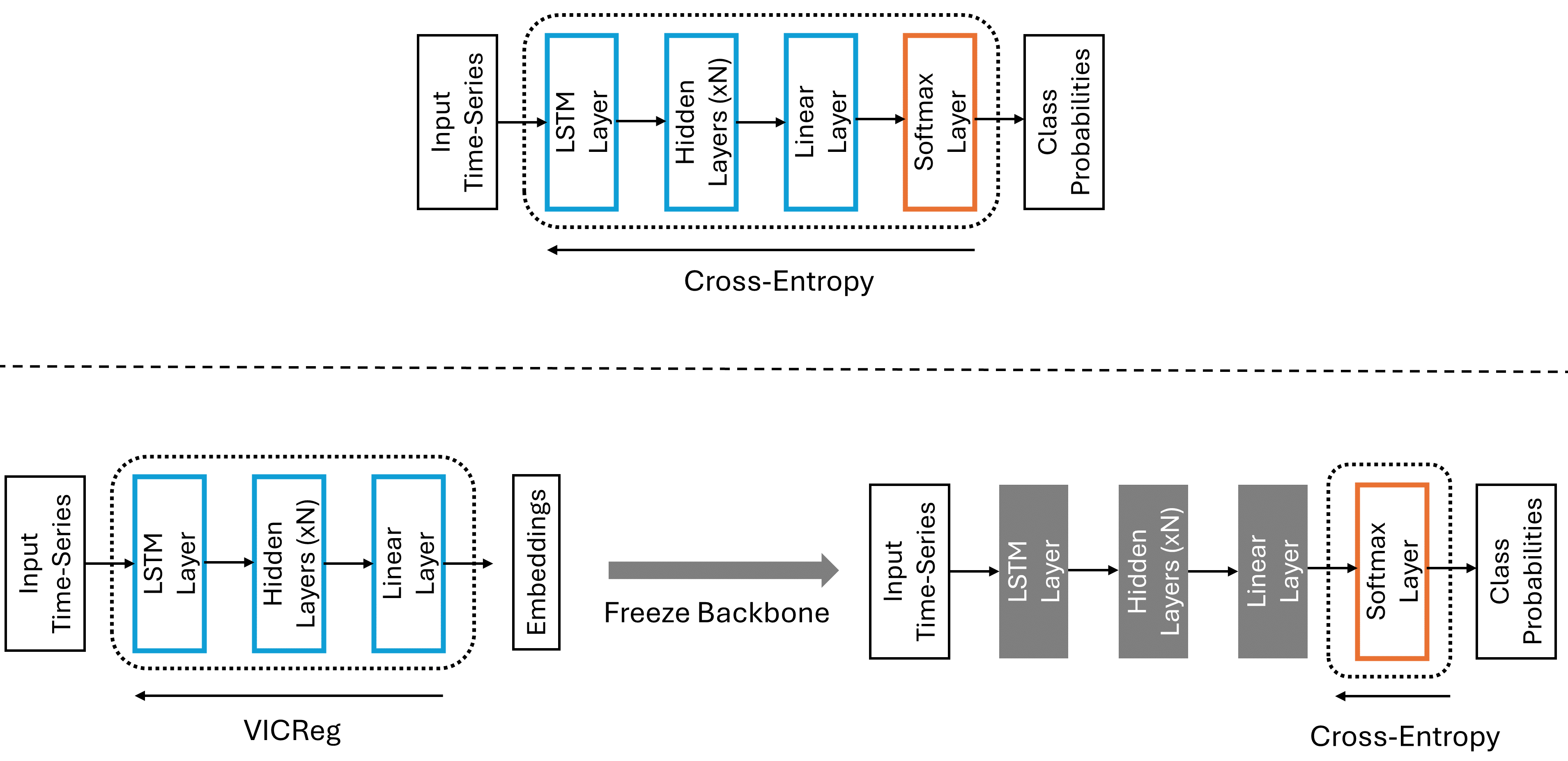}
    \caption{The deep network architecture comprises a backbone (blue boxes) and a linear softmax classification head (orange boxes). In conventional supervised training (top), both are optimized jointly using XEnt loss. In the proposed method (bottom), the backbone is first pre-trained with VICReg and then frozen. The classification head is trained on top using XEnt loss \cite{raghu_self-supervised_2024}.}
    \label{fig:lstm_architecture}
\end{figure*}

The architecture of the LSTM employed in this work consisted of a `backbone' and a `classification head' (Figure \ref{fig:lstm_architecture}). The backbone consisted of an LSTM layer with $128$ units followed by $2$ hidden layers each consisting of $128$ units with layer normalization and ReLU activation. The classification head consisted of a single linear layer with softmax activation. 

For training LDA-R and LSTM-R, all five trials of ramp data were used, with one random ramp trial serving as validation data for LSTM-R. For training LSTM-D and LSTM-V, one random continuous dynamic trial was reserved for validation, and four of the remaining five trials were used for training. LDA-D was trained using all five continuous dynamic trials. An additional trial, separate from the validation and training sets, was collected but excluded from the main analysis. This held-out `test trial' was reserved to verify the integrity of the training process and for potential future offline analyses.

The LSTM-R and LSTM-D models were trained end-to-end using XEnt loss with labeled ramp data and CRT-labeled continuous dynamic training trials, respectively. In contrast, the LSTM-V model followed a two-stage training process. First, the backbone was pre-trained on unlabeled continuous dynamic training trials using the VICReg loss. During this pre-training phase, three types of augmentations were applied to each sample to enhance the model's robustness and generalization capabilities: 1) random lag of $\pm4$ frames, 2) random feature scaling ($\mu = 1, \sigma = 0.05$), and 3) corruption through additive white Gaussian noise ( $\mu = 0, \sigma = 0.05$). Subsequently, the pre-trained backbone was frozen, and the classification head was trained on top of it with XEnt loss using the CRT-labeled continuous dynamic training trials.

For all deep models, the batch size was set to $256$ and AdamW \cite{loshchilov_decoupled_2019} optimizer with a weight decay value of $\SI{1e-3}{}$ was used to optimize the model parameters. The learning rate was set to $\SI{1e-3}{}$ when training the backbone and classification head separately, and was set to $\SI{1e-4}{}$ when training end-to-end with XEnt loss. Early stopping was employed to stop the training after $10$ epochs of no improvement in validation loss. All features were standardized to have zero mean and unit standard deviation based on the corresponding training data in all cases.

\subsubsection{Post-Processing}
This study employed confidence-based rejection, a technique previously shown to improve the usability of myoelectric control systems \cite{scheme_confidence-based_2013, robertson_effects_2019}. This approach rejects predictions with low confidence scores and thereby helps to mitigate the impact of noisy or misclassified data. Analysis conducted in our previous work \cite{raghu_self-supervised_2024} was used to inform the selection of a rejection threshold. Total error rates (classification error including decisions that were incorrectly classified as `No Movement') were compared for different rejection thresholds, and results indicated that the total error rate started to increase at a threshold of about $\mathrm{Th}=0.5$ for all classifiers. Since an increase in total error rate indicates over-rejection, which can lead to decreased responsiveness and user frustration in a myoelectric control system, this metric provided a way to select a threshold that balanced error mitigation and over-rejection.

\subsubsection{Control}

A proportional velocity control scheme was used, similar to man y previous works \cite{scheme_confidence-based_2013, scheme_motion_2014}, enabling participants to modulate cursor speed based on contraction intensity during the usability test. The proportional control (PC) value was derived for each of the classifiers independently using a method inspired by \cite{scheme_motion_2014}, as follows:

\begin{enumerate}
    \item \textbf{Training PC values}: The MAV of the EMG signal was summed across all channels for each frame in the training data to create a training PC value per frame.    
    \item \textbf{Class-specific normalization}: The distribution of training PC values was established for each class based on the classifier predicted labels of the frames in the training data. The minimum and maximum values for each class were then used to inform class-specific normalization coefficients. 
    \item \textbf{Normalization during inference}: During real-time operation, the MAV values were summed across the channels and normalized by the class-specific coefficients for the associated classifier output. 
    \item \textbf{Sigmoid transfer function}: Finally, the normalized test PC value was passed through a Sigmoid function ($f(x) = \sigma(a(x - x_0))$) to create the final proportional control value. This step ensures a smooth and bounded output suitable for cursor velocity control.
\end{enumerate}

The \nth{10} and \nth{95} percentiles were chosen for class-specific normalization to mitigate the effects of outliers, as informed by pilot studies. Similarly, the parameters $a, x_0$ for the Sigmoid transfer function $f(x) = \sigma(a(x - x_0))$ were set as $a = 10, x_0=0.5$ based on pilot studies. These values were selected to achieve a balance between responsiveness for larger movements and the ability to perform fine control for smaller adjustments, by mitigating the impact of minor variations in contraction intensity.

\subsection{3DoF ISO Fitts Law Test}

\begin{figure}[htbp]
    \centering
    \begin{subfigure}[b]{0.9\linewidth}
         \centering
         \includegraphics[width=\textwidth]{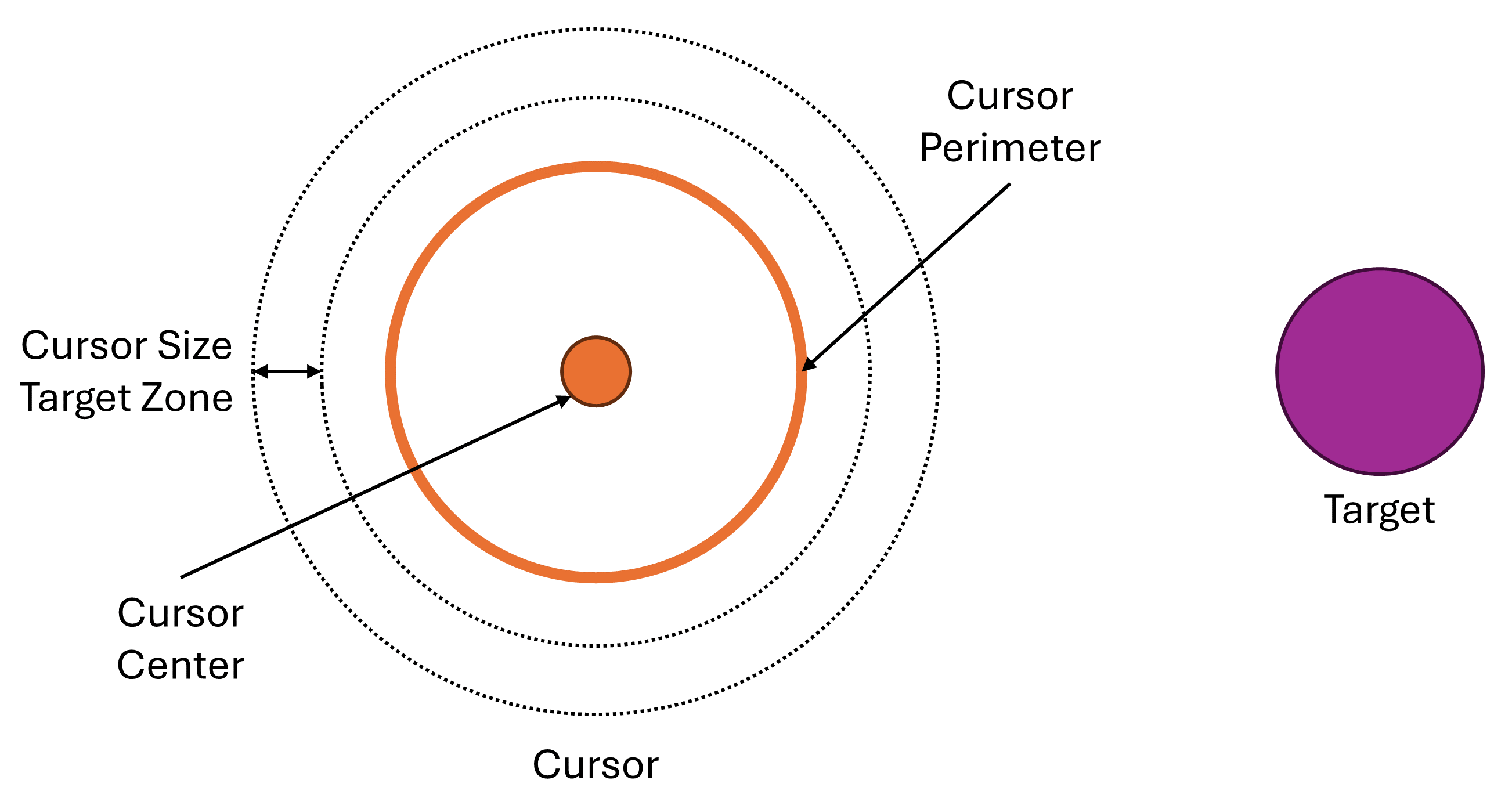}
         \caption{}
         \label{fig:fitts_env}
     \end{subfigure}

     \begin{subfigure}[b]{0.8\linewidth}
         \centering
         \includegraphics[width=0.5\textwidth]{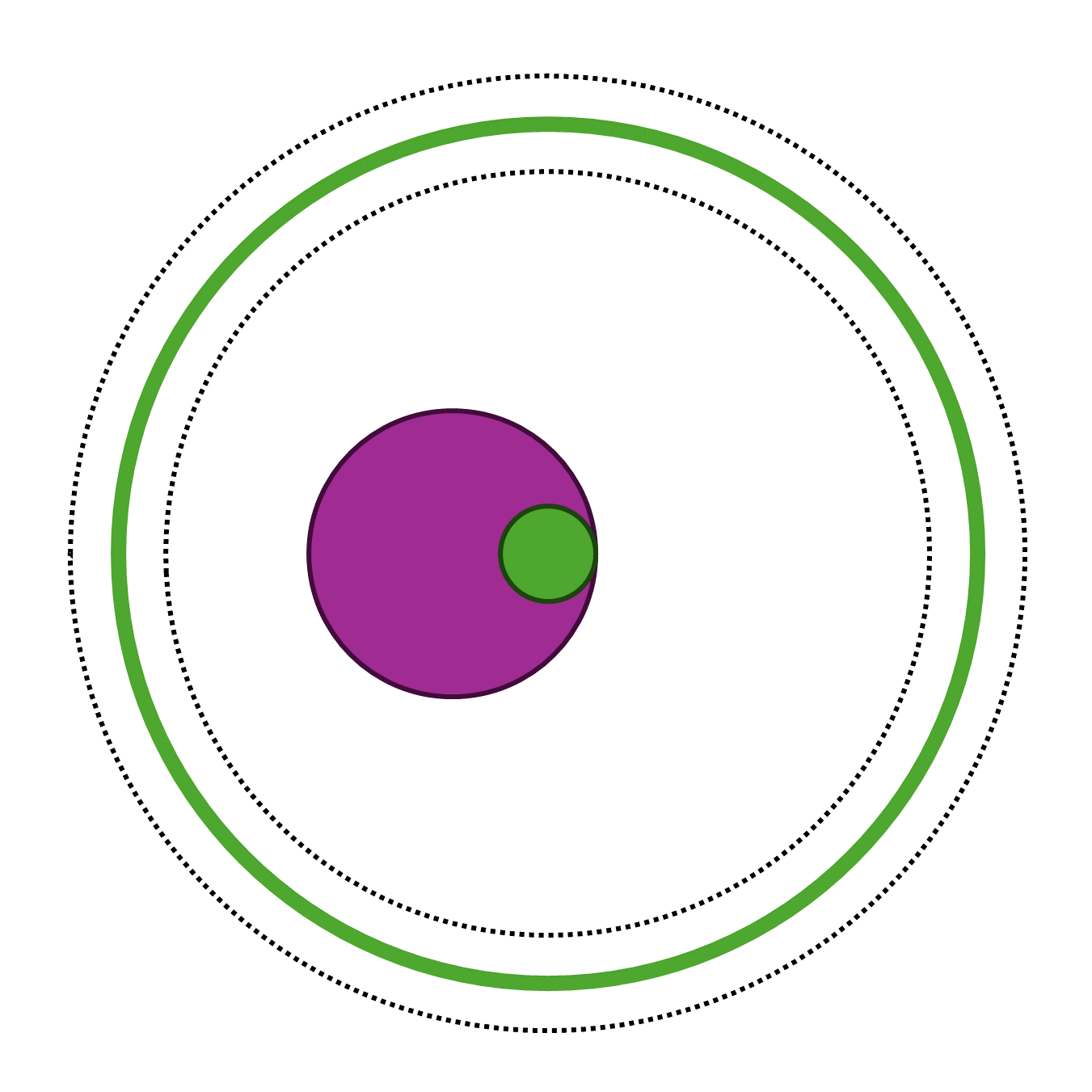}
         \caption{}
         \label{fig:fitts_feedback}
     \end{subfigure}     
    \caption{The 3DoF Fitts' law test employed in this work. (a) The purple filled circle indicates the target in the XY direction, and the dashed concentric circles around the center represents the target zone for the cursor size. The participants were instructed to move the center of the orange cursor into the target area, and to get the cursor size (perimeter) in the target zone by shrinking or expanding the cursor. (b) Visual feedback was provided when the users achieved one or more task criteria. The cursor perimeter turned green to indicate the target size was achieved, and the cursor center turned green to indicate that the center was inside the target. Note that this feedback was independent, meaning the element turned green when the associated objective was achieved (not shown).}
    \label{fig:fitts}    
\end{figure}

This section describes the 3DoF Fitts' Law test (hereafter referred to as the Fitts' Law test) used to evaluate the performance of the different classification algorithms. While adhering to the core principles of the ISO 9241-9 standard commonly used in Human-Computer Interaction (HCI) research \cite{soukoreff_towards_2004}, the virtual environment was extended to incorporate a 3D target generation scheme that is subsequently projected onto a 2D plane for display, similar to the approach employed in \cite{smith_real-time_2014}. Targets were generated on the surface of a virtual sphere, and their positions were projected onto the 2D plane to determine the target location on the XY plane. The `depth' or `z' component of each target's 3D position was utilized to control the desired size of the cursor at that target. A visual representation of the environment is presented in Figure~\ref{fig:fitts}.

\subsubsection{Participant Control and Visual Cues}

Participants controlled a cursor on the screen with two independent attributes (Figure~\ref{fig:fitts_env}):
\begin{itemize}
    \item \textbf{Location}: Represented by a dot, indicating the cursor's location within the 2D plane (X and Y coordinates).
    \item \textbf{Size}: Represented by a hollow circle surrounding the cursor center. This visual cue indicated the size of the cursor's outer perimeter, and traveled with the location cursor.
\end{itemize}

Specific muscle contractions were mapped to cursor control as follows:
\begin{itemize}
    \item \textbf{WF / WE}: moved the cursor in the positive/negative X-direction, respectively.
    \item \textbf{WP / WS}: Moved the cursor in the positive/negative Y-direction, respectively.
    \item \textbf{HO / HC} Increased/decreased the diameter (equivalently, the perimeter) of the hollow circle surrounding the cursor center, respectively.
\end{itemize}

The proportional control value controlled the speed of all three cursor dynamics: a higher control value resulted in a faster movement in the corresponding direction (X/Y) or a faster increase/decrease in cursor diameter, depending on the classifier output.  To ensure fair comparison across different cursor attributes and minimize potential biases related to movement speed, the speed was normalized such that it took the same duration to traverse the expected range of each cursor attribute (X, Y, and diameter) within the Fitts' Law task space.

\subsubsection{Target Presentation and Task Execution}

The virtual environment presented participants with a series of circular targets, each displayed one at a time to minimize cognitive load. Similar to conventional ISO Fitts' Law tests, participants navigated the cursor back and forth between targets on the virtual sphere. Critically, a successful target acquisition required meeting two criteria simultaneously for a dwell time of \SI{3}{\second}:

\begin{enumerate}
    \item \textbf{Cursor Center Position}: The center dot of the cursor had to be located within the designated target area.
    \item \textbf{Cursor Size}: The perimeter of the hollow circle surrounding the cursor center had to match the size requirement indicated by the concentric circles displayed around the target itself (shown in Figure~\ref{fig:fitts_env}).
\end{enumerate}

The environment provided real-time visual feedback to the user (Figure \ref{fig:fitts_feedback}). For example, the cursor's perimeter would turn green when it reached the appropriate size within the target zone. If either of these criteria was not met continuously for the entire \SI{3}{\second} dwell time window, the attempt was considered unsuccessful. Participants were allowed a maximum of \SI{13}{\second} to achieve and maintain both criteria for a successful target acquisition. The \SI{3}{\second} dwell time window was chosen to assess the user's ability to remain still in a desired position, deemed to be crucial for precise control in real-world applications. The \SI{13}{\second} maximum acquisition time, determined through pilot studies, provided sufficient opportunity for users to achieve and maintain both criteria while mitigating the risk of frustration and fatigue during the experiment.

It is worth noting that once the test began, the cursor was \textbf{not} modified in any way (e.g., reset to the center of the screen); the user immediately moved from one target to the next until the end of the test.

\subsubsection{Evaluation}

This section describes the evaluation approach used to assess user performance in the Fitts' Law test. Participants were presented with a total of 26 targets for each classifier. The first 13 trials were excluded from the analysis, as participants were encouraged to use them to familiarize themselves with the specific classifier and its corresponding proportional control. This approach aimed to ensure that the analyzed data reflected performance after participants had gained some experience with the system. All final 13 attempts were included in the evaluation, regardless of whether the target was successfully acquired within the allotted time or not. For unsuccessful trials (those exceeding the time limit), a movement time of \SI{13}{\second} (the maximum allowed time) was assigned. This inclusive approach ensures that the analysis captures the full range of user performance, including potential difficulties or challenges encountered during the task. 

A $5\times10$ balanced Latin square design was employed to counterbalance the order of classifier presentation across participants. The Latin square algorithm and order are delineated in \cite{sheehe_latin_1961}. In this approach, each classifier appears an equal number of times in each position within the test sequence. This controls for any systematic bias that might arise due to the order in which participants encounter the different classifiers.

This study employed six established online metrics \cite{robertson_effects_2019} along with a novel metric to evaluate user performance:
\begin{enumerate}
    \item Completion Rate: Reflects the percentage of targets successfully acquired within the allotted time.
    \item Movement Time: Represents the average time required to acquire a single target.
    \item Throughput (bits/s): Measures the rate of information, defined as the ratio of Index of Difficulty (ID) and movement time.
    \item Path Efficiency: Represents the ratio of the ideal path length to the actual movement path taken by the user. A value closer to 1 indicates a more efficient movement path.
    \item Stopping Distance: Measures the amount of cursor movement during the dwell time, reflecting the user's ability to maintain a stable cursor position once the target is reached.
    \item Overshoots: Counts the number of times the user's cursor size exceeded the target zone boundaries during an attempt, or the number of times the cursor was inside the target and then moved out, indicating challenges in precision control.
    \item Instability: A novel metric quantifying cursor jitter during movement. Similar to the offline instability metric \cite{tallam_puranam_raghu_analyzing_2022, raghu_decision-change_2023}, it is calculated as the average number of consecutive decisions that differ from one another in the decision stream, excluding blips to the ``No Movement'' class. Lower values indicate smoother and more stable control, with ideal values ranging from 0 to 2, reflecting the minimum number of transitions required between DoFs to reach the target.
\end{enumerate}

The Fitts' Law test employed a fixed ID of $\approx3.09$ bits (Amplitude = $300$ pixels, Width = $40$ pixels), a value determined through pilot studies, to balance test duration and user fatigue while ensuring sufficient sensitivity to capture meaningful variations in user performance. Manhattan distance, rather than Euclidean distance, was used to calculate the metrics. This choice acknowledges the inherently sequential nature of classification-based control.

Analysis focused on assessing whether any of the non-baseline models exhibited a significant improvement in the online metrics as compared to the baseline LDA-R model. A one-factor repeated-measures ANOVA with a significance level of $\alpha = 0.05$ was conducted to assess the effects of classifier on user performance metrics. For metrics exhibiting statistically significant effects ($p < 0.05$), post-hoc comparisons with Benjamini–Yekutieli correction were conducted to compare the performance of each non-baseline classifier to the baseline LDA-R model. Additionally, Cohen's d effect sizes were calculated for metrics with significant effects to quantify the magnitude of these differences. Effect sizes were interpreted according to the following categories from \cite{sawilowsky_new_2009}:  $d = 0.01$ = Very  Small,  $d=0.2$  =  Small,  $d =0.5$  =  Medium,  $d =0.8$ =  Large,  $d =1.2$  =  Very  Large,  and  $d = 2.0$  =  Huge.

\subsubsection{User Experience}

In addition to the online performance metrics, this study incorporated a user experience survey to gather subjective feedback from participants. Following the completion of the Fitts' Law test with each classifier, participants were asked to rate their overall experience and their satisfaction with cursor control using each of the six control interfaces: up, down, left, right, expansion, and contraction. Ratings were provided on a 5-point Likert scale (1: Poor, 5: Excellent). Average scores will be reported to provide insights into the user experience with different classifiers. Additionally, participants were encouraged to provide qualitative comments regarding their experience with each classifier, highlighting any notable strengths, weaknesses, or specific issues encountered.

\begin{figure*}[b]
    \centering        
    \includegraphics[width=0.9\textwidth]{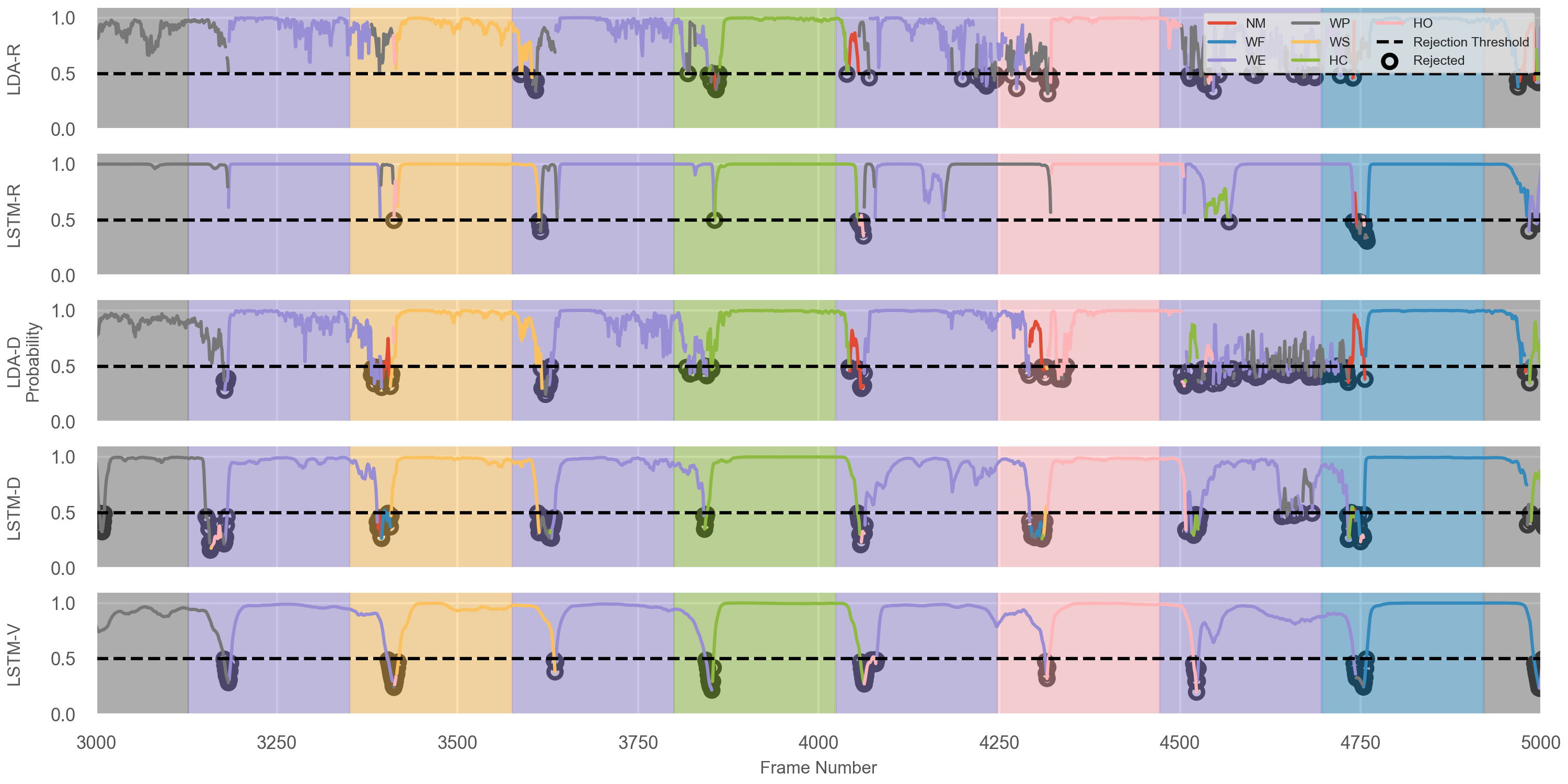} 
    \caption{Decision streams obtained from the different classifiers for the held out test set from an example participant. Colours of the data points represent the classifiers decision (as indicated in the legend), and values represent confidence; shaded regions denote the prompted class. Classifiers denoted by -R were trained with Ramp data, -D were trained with continuous dynamic data, and -V denotes the VICReg pre-trained model with the continuous dynamic data. The dashed horizontal lines indicate the rejection threshold of $0.5$ used in this study, and rejected decisions are marked with black circles.} 
    \label{fig:decision_stream}
\end{figure*}

\section{Results}

Figure \ref{fig:decision_stream} provides a visual comparison of the decision streams generated by the different classifiers for a representative offline test trial using continuous dynamic data. The line colors and shaded background colors denote the classifier outputs and user prompts, respectively. These decision streams, derived from a held-out test set, offer insights into the models' behavior during steady-state and transitions, providing context for interpreting the subsequent quantitative performance metrics. For instance, the figure suggests that LSTMs, in general, exhibit fewer fluctuations during steady-state and transitions compared to LDA, with LSTM-V demonstrating the least fluctuations overall. Moreover, when confidence-based rejection is applied, this effect is further amplified, particularly for LSTM-V and, to a lesser extent, LSTM-D. The reduction in erroneous predictions and smoother transitions observed with rejection in these models hints at their potential advantages in online performance.

\begin{figure*}[htbp]
    \centering    
        \includegraphics[width=0.7\textwidth]{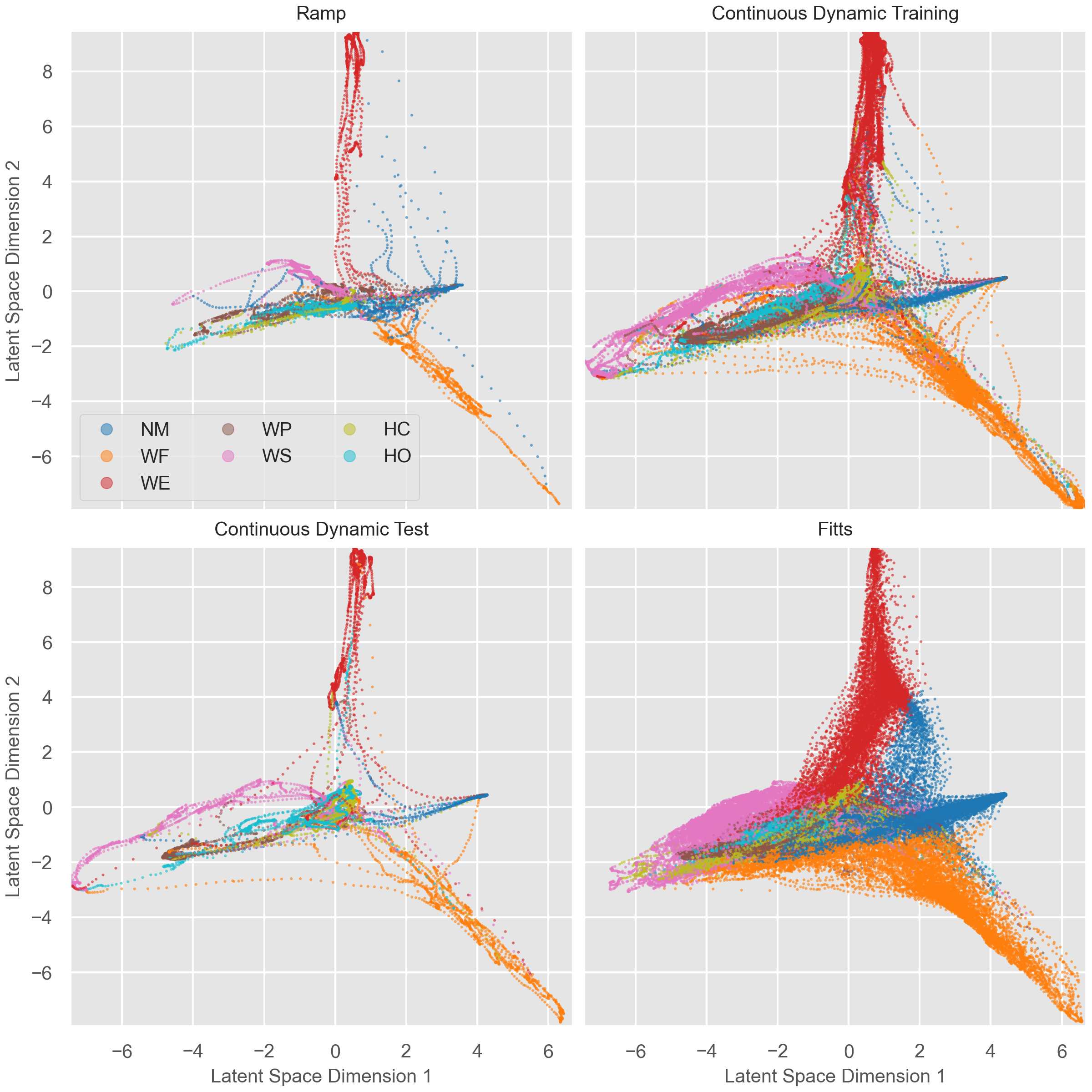}
    \caption{A 2D PCA projection of the VICReg latent space embeddings obtained from training, test, and Fitts data from an example participant. Top left: ramp training data (5 reps), top right: continuous dynamic training data (4 trial), bottom left: continuous dynamic test data (1 trial), bottom right: Data from Fitts test (all 5 models). Colors for the continuous dynamic training/test, and ramp training data indicate \emph{labels}, and colors for the Fitts data indicate \emph{predictions by LSTM-V}.}
    \label{fig:latent_space}
\end{figure*}

Figure~\ref{fig:latent_space} presents 2-dimensional visualizations of the latent space learned by the VICReg-trained backbone model using Principal Component Analysis (PCA). This visualization reveals a clustered structure in the latent space, with data points corresponding to each class appearing grouped despite the absence of explicit class labels during pre-training. Furthermore, inter-class transitions in the continuous dynamic data appear to be encoded as strand-like structures connecting these clusters. In contrast, the ramp data, which lacks such transitions due to the protocol design, exhibits distinct clusters without these connecting strands (except for the NM class). Interestingly, the data collected during the Fitts' Law test (bottom right) appears to exhibit a similar clustered structure with connecting strands.

\begin{figure*}[htbp]
    \centering    
        \includegraphics[width=0.9\textwidth]{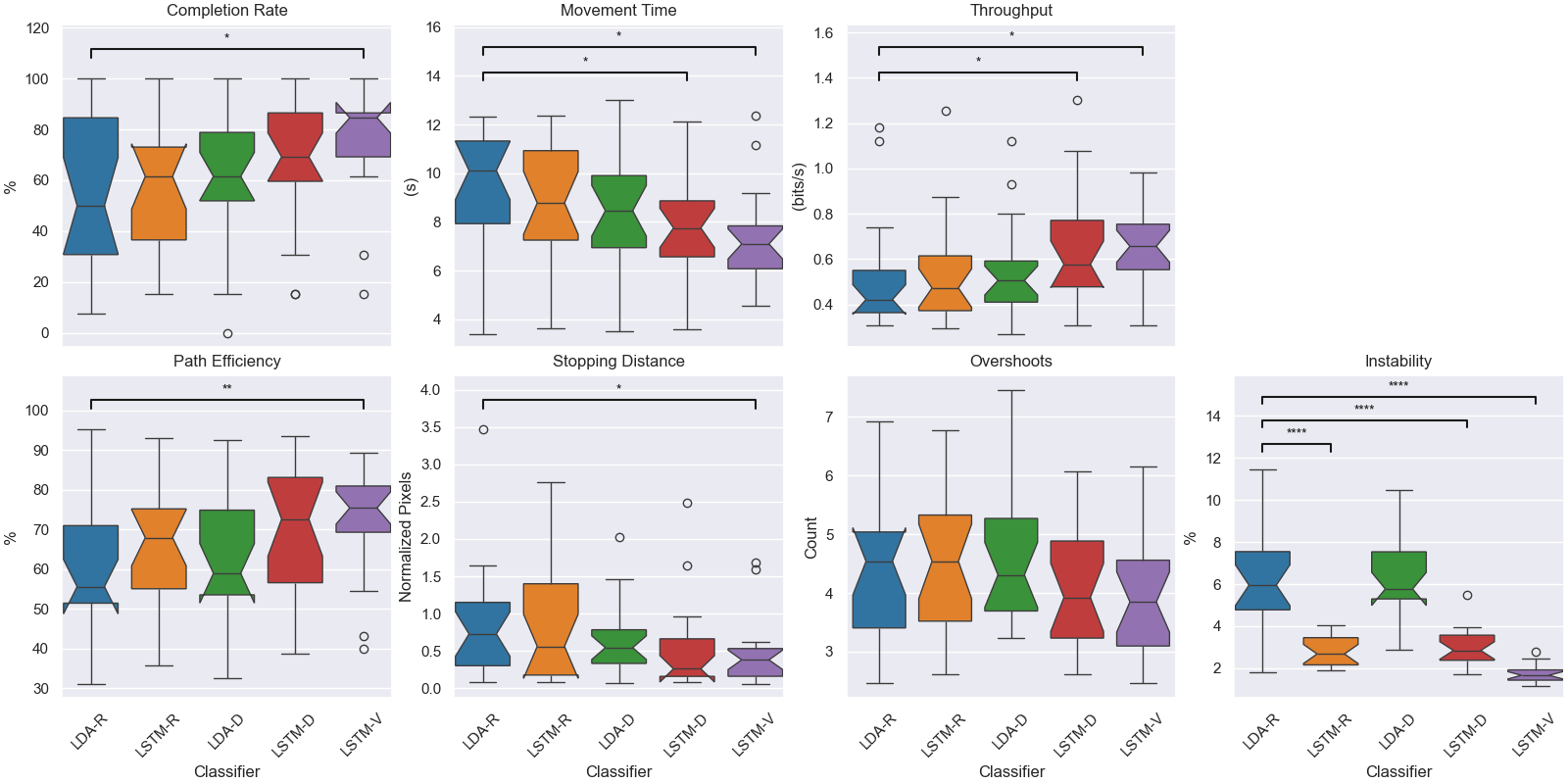}
    \caption{Performances of the different classifiers across subjects. Classifiers denoted by -R were trained with Ramp data, -D were trained with continuous dynamic data, and -V denotes the VICReg pre-trained model with the continuous dynamic data. Only significant differences are highlighted for clarity: *=$p<0.5$, **=$p< 0.01$, ****=$p< 0.0001$.}
    \label{fig:metrics}
\end{figure*}

Figure \ref{fig:metrics} presents box plots comparing the seven Fitts metrics across the classifiers. The ANOVA results revealed significant differences in all metrics ($p < 0.047$ in all cases) except for stopping distance.

Post-hoc testing revealed that LSTM-V had the best completion rate, and significantly outperformed LDA-R with a Large effect size ($p < 0.02$; $d > 0.8$). No other significant differences emerged in completion rate. A similar trend emerged for path efficiency where LSTM-V significantly outperformed the LDA-R with a Medium effect size ($p < 0.005$; $d > 0.78$). No other significant differences emerged in path efficiency.

For movement time, LSTM-V again emerged as the best model, exhibiting the smallest movement time, and significantly outperformed LDA-R with a Large effect size ($p < 0.001$; $d > 0.8$). LSTM-D also exhibited a significant reduction in movement time compared to LDA-R with a Medium effect size ($p < 0.03$; $d > 0.5$). No other significant differences emerged in movement time. A similar trend emerged for throughput where LSTM-V significantly outperformed the LDA-R with a Medium effect size ($p < 0.05$; $d > 0.6$). LSTM-D also exhibited a significant increase in throughput compared to LDA-R with a Medium effect size ($p < 0.02$; $d > 0.5$). No other significant differences emerged in throughput.  

For instability, a significant improvement over LDA-R was observed across all temporal models (LSTM-R, LSTM-D, and LSTM-V), with each demonstrating at least a Very Large effect size ($p < 0.0001, d > 1.8$). LSTM-V performed the best, exhibiting the largest effect size ($d>2.6$) across all the models. No significant difference emerged between LDA-R and LDA-D.

Finally, for overshoots, although the ANOVA results indicated significant differences overall, post-hoc comparisons did not reveal any significant differences between individual classifier pairs.

\begin{figure*}[htb]
    \centering    
        \includegraphics[width=0.8\textwidth]{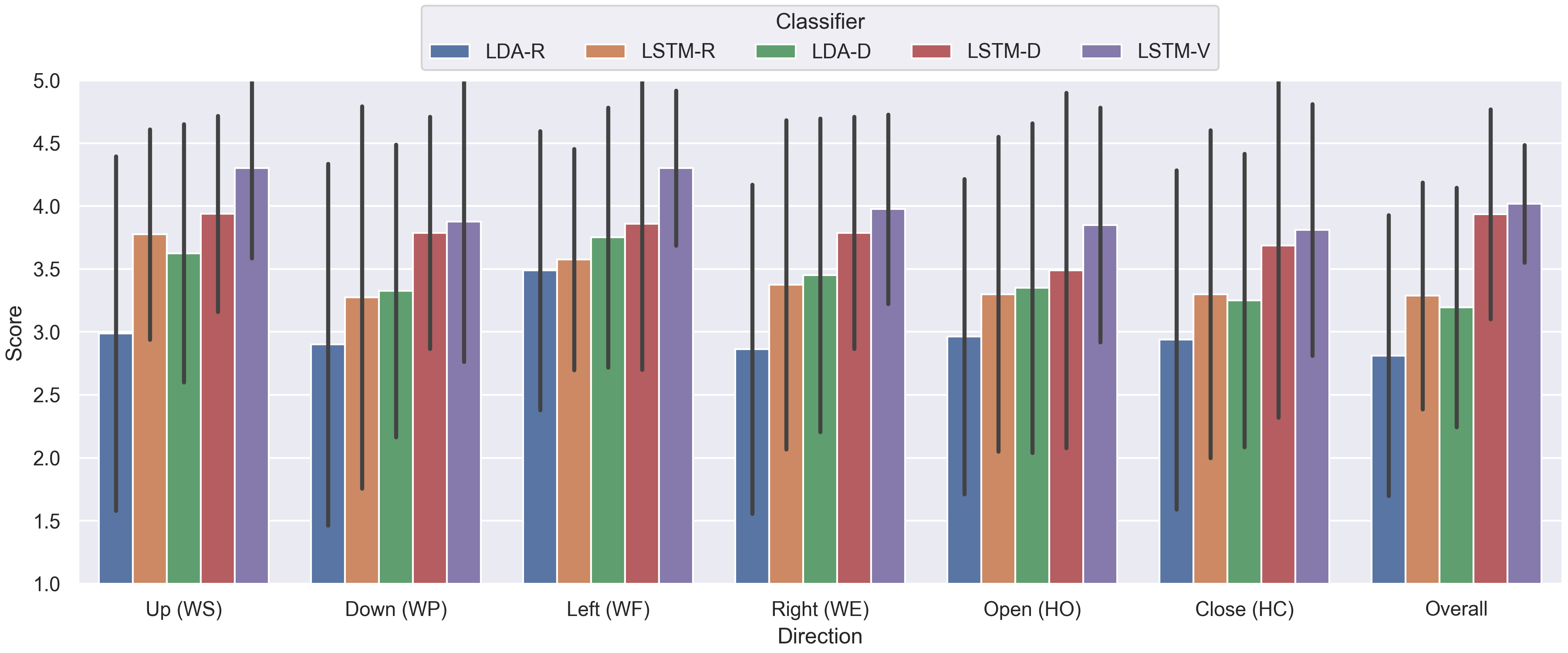}
    \caption{Participant ratings for each class across the different classifiers used in the Fitts task. Labels indicate the control interface and the corresponding class associated with the interface. Note that overall ratings were also provided by the participant, and \textbf{not} the average of the scores across the different interfaces.}
    \label{fig:likert_scores}
\end{figure*}

Figure~\ref{fig:likert_scores} depicts a bar plot comparing the average ratings provided by the participants for all the models across the different control interfaces. Consistent with the online performance metrics, LSTM-V consistently achieved the highest user ratings, outperforming all other models across all control interfaces. In contrast, LDA-R consistently received the lowest user ratings across all control interfaces. LSTM-D, while not reaching the same level of user satisfaction as LSTM-V, consistently outperformed LDA-R. Interestingly, LSTM-R received user ratings comparable to LDA-D, and slightly more favourable ratings compared to LDA-R.

In summary, the online performance metrics and subjective user ratings consistently highlighted the superiority of the LSTM-V model. The LDA-R model, while serving as a valuable baseline, exhibited lower performance across most metrics. These results suggest that training with continuous dynamic data and leveraging SSL techniques like VICReg can lead to substantial improvements in myoelectric control performance and user experience.

\section{Discussion}

Our results demonstrate that temporal models, specifically LSTM-D and LSTM-V trained on continuous dynamic data, achieved superior performance in the Fitts' Law test compared to the baseline LDA-R model. Specifically, the LSTM-V significantly outperformed the LDA-R model across virtually all performance metrics. It is interesting to note that significant improvements in movement time and throughput arose despite the use of optimistic movement times for incomplete trials. This approach (wherein failed trials were assigned a completion time equal to that of the timeout, despite not actually being completed) was adopted to enable the inclusion of those trials in the analysis of all metrics. However, this effectively introduces an interaction wherein algorithms with lower completion rates enjoy somewhat inflated throughput. Consequently, because the LSTM-V also enjoyed significantly better completion rates than LDA-R, the true differences in movement time and throughput are even larger than reported in \ref{fig:metrics}. 

These finding align with our previous offline results, which indicated that GRUs and LSTMs trained with continuous dynamic data showed significant improvements over LDA-R. This suggests that temporal models, such as LSTMs, are particularly adept at harnessing the richer information present in continuous dynamic data, resulting in improved online control. Interestingly, while LDA-D also leveraged continuous dynamic data, its improvements over LDA-R were not statistically significant, further highlighting the unique capabilities of temporal models in this context. 

The clustered structure with connecting strands observed in the latent space (Figure~\ref{fig:latent_space}) offers insights into the superior performance of LSTM-V and LSTM-D. The Fitts' Law data displayed a remarkably similar structure to that of the continuous dynamic data. This observation suggests that during closed-loop myoelectric control tasks, users naturally produce a diverse range of muscle contractions, encompassing transitions between different classes (as exemplified by the red (WE) trajectories leading to the pink (WS) cluster in Figure~\ref{fig:latent_space}).  Therefore, it is reasonable to expect that models trained on the dynamics captured in continuous data would outperform those trained solely on ramp data in online settings. Furthermore, this observation strongly supports the hypothesis that the latent space learned from continuous dynamic data may generalize better to real-world usage, and our results warrant further investigation to explore the full extent of this generalization capability. 

The observation in Figure \ref{fig:decision_stream} that LSTM-V exhibits a larger classifier confidence range than other models, particularly for incorrect decisions and during transitions, is also noteworthy. This behavior aligns with our previous offline results \cite{raghu_self-supervised_2024}, and with recent findings in the broader literature suggesting that self-supervised learning can improve uncertainty estimation \cite{NEURIPS2019_a2b15837}. The ability of LSTM-V to express lower confidence in uncertain situations could contribute to its superior performance, especially when combined with confidence-based rejection.

Participant ratings also support this trend, revealing a clear preference for temporal models over LDA. LSTM-V and LSTM-D were rated the best models in the test, approximately $1$ point higher overall than the others. Interestingly, although LDA-D was trained on continuous dynamic data, the ratings were comparable or slightly lower than LSTM-R. These results indicate that continuous training data may only provide benefit when combined with temporal models that can leverage them fully. Furthermore, the magnitude of this preference for temporal models, as reflected in the ratings, appears to outweigh the differences observed in throughput, suggesting that factors beyond pure performance metrics contribute to user satisfaction. It is also worth noting that ratings across different degrees of freedom were relatively close, though WF and WS received slightly higher ratings than other classes.

Our internal pilot testing revealed a small, but not significant, advantage for LSTMs over GRUs in the backbone architecture, but not for feature sets beyond the LSF4 features employed in this study. This suggests that deep temporal models, with their capacity to learn complex representations effectively from the input data, may be less sensitive to the specific choice of recurrent unit or handcrafted features, especially when trained with continuous dynamic data. This observation aligns with recent findings in discrete myoelectric control, which also highlight the ability of deep learning models to leverage temporal structure for robust performance \citeauthor{eddy_bigdata}. Nevertheless, a systematic evaluation of different feature sets and architectural choices in online scenarios would be valuable to confirm this hypothesis and explore potential interactions between these factors.

Beyond the quantitative ratings, qualitative feedback from participants revealed a strong correlation between the perceived `jitteriness' of the cursor and the online instability metric. Non-temporal models, exemplified by LDA-R, were frequently described as exhibiting excessive jitter, contributing to a less enjoyable user experience. This observation aligns with the quantitatively higher instability scores recorded for these models.

Participants also reported encountering transition errors, particularly when transitioning to Wrist Extension (WE) or Hand Close/Hand Open (HC/HO) contractions. While the frequency of these errors could be influenced by the specific design of the Fitts' Law test, the qualitative feedback suggests that certain transition errors were more noticeable or disruptive to the user than others. These inconsistencies in performance across different movement transitions can significantly impact user satisfaction and hinder the overall usability of a myoelectric control system. Future research should investigate whether specific transitions are inherently more challenging than others and identify potential strategies for mitigating their impact.

The qualitative feedback from participants also highlights the importance of ensuring consistent performance across different contraction classes to enhance the overall user experience. While achieving exceptional performance for certain classes is valuable, it's equally important to ensure that the baseline performance across all classes is satisfactory. In essence, enhancing the overall user experience necessitates a focus on elevating the baseline performance across all classes, not solely maximizing peak performance on select classes. 

The observed average throughput for LDA-R in our 3DoF Fitts' Law test is consistent with previous findings \cite{smith_real-time_2014}. Notably, the researchers in \cite{smith_real-time_2014} also reported a consistent reduction in throughput with increasing DoF (approximately $50\%$ reduction per additional DoF), underscoring the inherent difficulty of 3DoF tasks. This challenge is further reflected in the relatively low completion rate observed for LDA-R in our study. However, the median completion rate of LSTM-V was $\approx1.7\mathrm{x}$ that of LDA-R ($84.6\%$ vs$50\%$), demonstrating that there is still significant room to be gained through improvements in sEMG classifier design. Future research should explore how such improvements translate to applications such as prosthesis or exoskeleton control.

In the Fitts' Law test, users not only controlled the direction of the cursor, but also the speed. In pilot testing, our initial approach to determining the parameters for proportional control relied on a method established for ramp data \cite{scheme_motion_2014}. However, continued study revealed limitations in directly transferring this scheme to the continuous dynamic data case. To address these limitations and enhance user experience across both data types, we made adjustments to thresholds and incorporated a sigmoid transfer function. While not the main focus of this article, this suggests that there may be benefits to be gained by exploring alternative proportional control schemes tailored for continuous dynamic data in myoelectric control. 

Similarly, individual user preferences were noted regarding speed and sensitivity during the experiment. This subjectivity may motivate the ability to personalize both classification and proportional control sensitivity parameters, such as thresholds, gains, rejection thresholds, and transfer function characteristics. Although we opted for fixed proportional control mappings and rejection thresholds in this work to ensure a fair comparison across the classifiers, future research in myoelectric control should consider integrating mechanisms for user-specific adjustments, acknowledging the inherent variability in how individuals experience and exert muscle contractions. Deep learning models, with their ability to learn complex relationships from data, might be particularly well-suited to this task. By simultaneously learning classification and user-specific preferences for proportional control, these models could potentially lead to more intuitive and adaptable control interfaces for myoelectric prostheses \cite{campbell_ciil}.

Admittedly, our study employed continuous dynamic training data trials that took moderately longer to collect than the ramp data trials. Although this was clearly beneficial to performance, it remains to be explored how best to reduce this training burden while retaining its benefits. The required number and composition of trials should be further studied, as should cross-user techniques like transfer learning and domain adaptation \cite{cote-allard_deep_2019, zhang_multi-source_2023, xu_cross-user_2024} to potentially reduce user burden while maintaining the observed benefits. Additionally, the data augmentation techniques used in this work were chosen to match our prior work \cite{raghu_enabling_2024}. However, the literature currently lacks established best practices for augmentation strategies and hyperparameter settings in a myoelectric control setting. Therefore, a systematic exploration is warranted to evaluate the effectiveness of these and other augmentation techniques specifically for online myoelectric control tasks. 

Given the observed benefits of incorporating added dynamics in this work, a promising avenue for future work involves incorporating a more diverse range of dynamic data into the training process, such as data obtained during closed-loop usage. Given its self-supervised nature and ability to operate without labeled data, VICReg is well-suited to harnessing the richness of such unlabeled dynamic contractions. By training on a more comprehensive set of real-world dynamic patterns, the model could potentially learn more robust representations that generalize better to online control.

\section{Conclusion}
In conclusion, this study evaluated the performance of various myoelectric control algorithms using a challenging 3DoF Fitts' Law task with continuous dynamic and ramp data. Our results demonstrate that training with continuous dynamic data, particularly when leveraging deep temporal models, leads to improved online usability. Furthermore, the VICReg SSL approach further enhances performance when utilizing continuous dynamic data. These findings open avenues for future research, including exploring the impact of incorporating additional movement dynamics during training, developing improved labeling strategies for continuous dynamic data, leveraging transfer learning for enhanced generalization with limited data, systematically analyzing the impact of various augmentations for sEMG-PR, and refining proportional control mapping with user preferences in mind.

The visualization of the latent space learned by VICReg suggests that continuous dynamic data better reflects data seen during online control. These findings support the potential of using continuous dynamic data for training myoelectric classifiers and highlight the promise of self-supervised approaches like VICReg in this domain.

While our study yielded promising results, it also leaves areas for future investigation. Future work should focus on developing techniques to reduce training data requirements and systematically evaluate data augmentation strategies tailored for online myoelectric control. By addressing these limitations and building upon our findings, we can advance the development of more intuitive, adaptable, and user-centric myoelectric control schemes.

\printbibliography

@article{cote-allard_deep_2019,
	title = {Deep {Learning} for {Electromyographic} {Hand} {Gesture} {Signal} {Classification} {Using} {Transfer} {Learning}},
	volume = {27},
	issn = {1558-0210},
	url = {https://ieeexplore.ieee.org/abstract/document/8630679},
	doi = {10.1109/TNSRE.2019.2896269},
	number = {4},
	urldate = {2024-05-13},
	journal = {IEEE Transactions on Neural Systems and Rehabilitation Engineering},
	author = {Côté-Allard, Ulysse and Fall, Cheikh Latyr and Drouin, Alexandre and Campeau-Lecours, Alexandre and Gosselin, Clément and Glette, Kyrre and Laviolette, François and Gosselin, Benoit},
	month = apr,
	year = {2019},
	pages = {760--771},
}

@article{tallam_puranam_raghu_analyzing_2022,
	title = {Analyzing the impact of class transitions on the design of pattern recognition-based myoelectric control schemes},
	volume = {71},
	issn = {1746-8094},
	url = {https://www.sciencedirect.com/science/article/pii/S174680942100731X},
	doi = {10.1016/j.bspc.2021.103134},
	urldate = {2024-05-14},
	journal = {Biomedical Signal Processing and Control},
	author = {Raghu, Shriram Tallam Puranam and MacIsaac, Dawn and Scheme, Erik},
	month = jan,
	year = {2022},
	pages = {103134},
}

@article{raghu_decision-change_2023,
	title = {Decision-{Change} {Informed} {Rejection} {Improves} {Robustness} in {Pattern} {Recognition}-{Based} {Myoelectric} {Control}},
	volume = {27},
	issn = {2168-2208},
	url = {https://ieeexplore.ieee.org/abstract/document/10254242/},
	doi = {10.1109/JBHI.2023.3316599},
	number = {12},
	urldate = {2024-05-14},
	journal = {IEEE Journal of Biomedical and Health Informatics},
	author = {Raghu, Shriram Tallam Puranam and MacIsaac, Dawn and Scheme, Erik},
	month = dec,
	year = {2023},
	pages = {6051--6061},
}

@misc{bardes_vicreg:_2022,
	title = {{VICReg}: {Variance}-{Invariance}-{Covariance} {Regularization} for {Self}-{Supervised} {Learning}},
	shorttitle = {{VICReg}},
	url = {http://arxiv.org/abs/2105.04906},
	doi = {10.48550/arXiv.2105.04906},
	urldate = {2024-05-21},
	publisher = {arXiv},
	author = {Bardes, Adrien and Ponce, Jean and LeCun, Yann},
	month = jan,
	year = {2022},
	note = {arXiv:2105.04906 [cs]},
}

@article{simao_review_2019,
	title = {A {Review} on {Electromyography} {Decoding} and {Pattern} {Recognition} for {Human}-{Machine} {Interaction}},
	volume = {7},
	issn = {2169-3536},
	url = {https://ieeexplore.ieee.org/document/8672131},
	doi = {10.1109/ACCESS.2019.2906584},
	urldate = {2024-05-23},
	journal = {IEEE Access},
	author = {Simão, Miguel and Mendes, Nuno and Gibaru, Olivier and Neto, Pedro},
	year = {2019},
	pages = {39564--39582},
}

@article{samuel_intelligent_2019,
	title = {Intelligent {EMG} {Pattern} {Recognition} {Control} {Method} for {Upper}-{Limb} {Multifunctional} {Prostheses}: {Advances}, {Current} {Challenges}, and {Future} {Prospects}},
	volume = {7},
	issn = {2169-3536},
	shorttitle = {Intelligent {EMG} {Pattern} {Recognition} {Control} {Method} for {Upper}-{Limb} {Multifunctional} {Prostheses}},
	url = {https://ieeexplore.ieee.org/document/8612917},
	doi = {10.1109/ACCESS.2019.2891350},
	urldate = {2024-05-23},
	journal = {IEEE Access},
	author = {Samuel, Oluwarotimi Williams and Asogbon, Mojisola Grace and Geng, Yanjuan and Al-Timemy, Ali H. and Pirbhulal, Sandeep and Ji, Ning and Chen, Shixiong and Fang, Peng and Li, Guanglin},
	year = {2019},
	pages = {10150--10165},
}

@article{phinyomark_feature_2018,
	title = {Feature {Extraction} and {Selection} for {Myoelectric} {Control} {Based} on {Wearable} {EMG} {Sensors}},
	volume = {18},
	copyright = {http://creativecommons.org/licenses/by/3.0/},
	issn = {1424-8220},
	url = {https://www.mdpi.com/1424-8220/18/5/1615},
	doi = {10.3390/s18051615},
	language = {en},
	number = {5},
	urldate = {2024-05-26},
	journal = {Sensors},
	author = {Phinyomark, Angkoon and N. Khushaba, Rami and Scheme, Erik},
	month = may,
	year = {2018},
	pages = {1615},
}

@article{scheme_confidence-based_2013,
	title = {Confidence-{Based} {Rejection} for {Improved} {Pattern} {Recognition} {Myoelectric} {Control}},
	volume = {60},
	issn = {1558-2531},
	url = {https://ieeexplore.ieee.org/abstract/document/6409423},
	doi = {10.1109/TBME.2013.2238939},
	number = {6},
	urldate = {2024-05-26},
	journal = {IEEE Transactions on Biomedical Engineering},
	author = {Scheme, Erik J. and Hudgins, Bernard S. and Englehart, Kevin B.},
	month = jun,
	year = {2013},
	pages = {1563--1570},
}

@misc{chollet2015keras,
  title={Keras},
  author={Chollet, Fran\c{c}ois and others},
  year={2015},
  howpublished={\url{https://keras.io}},
}

@misc{tensorflow2015-whitepaper,
title={ {TensorFlow}: Large-Scale Machine Learning on Heterogeneous Systems},
url={https://www.tensorflow.org/},
note={Software available from tensorflow.org},
author={
    Mart\'{i}n~Abadi and
    Ashish~Agarwal and
    Paul~Barham and
    Eugene~Brevdo and
    Zhifeng~Chen and
    Craig~Citro and
    Greg~S.~Corrado and
    Andy~Davis and
    Jeffrey~Dean and
    Matthieu~Devin and
    Sanjay~Ghemawat and
    Ian~Goodfellow and
    Andrew~Harp and
    Geoffrey~Irving and
    Michael~Isard and
    Yangqing Jia and
    Rafal~Jozefowicz and
    Lukasz~Kaiser and
    Manjunath~Kudlur and
    Josh~Levenberg and
    Dandelion~Man\'{e} and
    Rajat~Monga and
    Sherry~Moore and
    Derek~Murray and
    Chris~Olah and
    Mike~Schuster and
    Jonathon~Shlens and
    Benoit~Steiner and
    Ilya~Sutskever and
    Kunal~Talwar and
    Paul~Tucker and
    Vincent~Vanhoucke and
    Vijay~Vasudevan and
    Fernanda~Vi\'{e}gas and
    Oriol~Vinyals and
    Pete~Warden and
    Martin~Wattenberg and
    Martin~Wicke and
    Yuan~Yu and
    Xiaoqiang~Zheng},
  year={2015},
}

@misc{loshchilov_decoupled_2019,
	title = {Decoupled {Weight} {Decay} {Regularization}},
	url = {http://arxiv.org/abs/1711.05101},
	doi = {10.48550/arXiv.1711.05101},
	urldate = {2024-06-02},
	publisher = {arXiv},
	author = {Loshchilov, Ilya and Hutter, Frank},
	month = jan,
	year = {2019},
	note = {arXiv:1711.05101 [cs, math]},
}

@article{harris_array_2020,
	title = {Array programming with {NumPy}},
	volume = {585},
	copyright = {2020 The Author(s)},
	issn = {1476-4687},
	url = {https://www.nature.com/articles/s41586-020-2649-2},
	doi = {10.1038/s41586-020-2649-2},
	language = {en},
	number = {7825},
	urldate = {2024-06-03},
	journal = {Nature},
	author = {Harris, Charles R. and Millman, K. Jarrod and van der Walt, Stéfan J. and Gommers, Ralf and Virtanen, Pauli and Cournapeau, David and Wieser, Eric and Taylor, Julian and Berg, Sebastian and Smith, Nathaniel J. and Kern, Robert and Picus, Matti and Hoyer, Stephan and van Kerkwijk, Marten H. and Brett, Matthew and Haldane, Allan and del Río, Jaime Fernández and Wiebe, Mark and Peterson, Pearu and Gérard-Marchant, Pierre and Sheppard, Kevin and Reddy, Tyler and Weckesser, Warren and Abbasi, Hameer and Gohlke, Christoph and Oliphant, Travis E.},
	month = sep,
	year = {2020},
	pages = {357--362},
}

@article{virtanen_scipy_2020,
	title = {{SciPy} 1.0: fundamental algorithms for scientific computing in {Python}},
	volume = {17},
	copyright = {2020 The Author(s)},
	issn = {1548-7105},
	shorttitle = {{SciPy} 1.0},
	url = {https://www.nature.com/articles/s41592-019-0686-2},
	doi = {10.1038/s41592-019-0686-2},
	language = {en},
	number = {3},
	urldate = {2024-06-03},
	journal = {Nature Methods},
	author = {Virtanen, Pauli and Gommers, Ralf and Oliphant, Travis E. and Haberland, Matt and Reddy, Tyler and Cournapeau, David and Burovski, Evgeni and Peterson, Pearu and Weckesser, Warren and Bright, Jonathan and van der Walt, Stéfan J. and Brett, Matthew and Wilson, Joshua and Millman, K. Jarrod and Mayorov, Nikolay and Nelson, Andrew R. J. and Jones, Eric and Kern, Robert and Larson, Eric and Carey, C. J. and Polat, İlhan and Feng, Yu and Moore, Eric W. and VanderPlas, Jake and Laxalde, Denis and Perktold, Josef and Cimrman, Robert and Henriksen, Ian and Quintero, E. A. and Harris, Charles R. and Archibald, Anne M. and Ribeiro, Antônio H. and Pedregosa, Fabian and van Mulbregt, Paul},
	month = mar,
	year = {2020},
	pages = {261--272},
}

@article{pedregosa_scikit-learn:_2011,
	title = {Scikit-learn: {Machine} {Learning} in {Python}},
	volume = {12},
	issn = {1533-7928},
	shorttitle = {Scikit-learn},
	url = {http://jmlr.org/papers/v12/pedregosa11a.html},
	number = {85},
	urldate = {2024-06-03},
	journal = {Journal of Machine Learning Research},
	author = {Pedregosa, Fabian and Varoquaux, Gaël and Gramfort, Alexandre and Michel, Vincent and Thirion, Bertrand and Grisel, Olivier and Blondel, Mathieu and Prettenhofer, Peter and Weiss, Ron and Dubourg, Vincent and Vanderplas, Jake and Passos, Alexandre and Cournapeau, David and Brucher, Matthieu and Perrot, Matthieu and Duchesnay, Édouard},
	year = {2011},
	pages = {2825--2830},
}

@article{waskom_seaborn:_2021,
	title = {seaborn: statistical data visualization},
	volume = {6},
	issn = {2475-9066},
	shorttitle = {seaborn},
	url = {https://joss.theoj.org/papers/10.21105/joss.03021},
	doi = {10.21105/joss.03021},
	language = {en},
	number = {60},
	urldate = {2024-06-03},
	journal = {Journal of Open Source Software},
	author = {Waskom, Michael L.},
	month = apr,
	year = {2021},
	pages = {3021},
}

@article{hunter_matplotlib:_2007,
	title = {Matplotlib: {A} {2D} {Graphics} {Environment}},
	volume = {9},
	issn = {1558-366X},
	shorttitle = {Matplotlib},
	url = {https://ieeexplore.ieee.org/document/4160265},
	doi = {10.1109/MCSE.2007.55},
	number = {3},
	urldate = {2024-06-03},
	journal = {Computing in Science \& Engineering},
	author = {Hunter, John D.},
	month = may,
	year = {2007},
	pages = {90--95},
}

@article{vallat_pingouin:_2018,
	title = {Pingouin: statistics in {Python}},
	volume = {3},
	issn = {2475-9066},
	shorttitle = {Pingouin},
	url = {https://joss.theoj.org/papers/10.21105/joss.01026},
	doi = {10.21105/joss.01026},
	language = {en},
	number = {31},
	urldate = {2024-06-03},
	journal = {Journal of Open Source Software},
	author = {Vallat, Raphael},
	month = nov,
	year = {2018},
	pages = {1026},
}

@article{mckinney_data_2010,
	title = {Data {Structures} for {Statistical} {Computing} in {Python}},
	url = {http://conference.scipy.org.s3-website-us-east-1.amazonaws.com/proceedings/scipy2010/mckinney.html},
	doi = {10.25080/Majora-92bf1922-00a},
	urldate = {2024-06-03},
	journal = {Proceedings of the 9th Python in Science Conference},
	author = {McKinney, Wes},
	year = {2010},
	pages = {56--61},
}

@article{sawilowsky_new_2009,
	title = {New {Effect} {Size} {Rules} of {Thumb}},
	volume = {8},
	copyright = {Copyright (c) 2009},
	issn = {1538-9472},
	url = {https://jmasm.com/index.php/jmasm/article/view/452},
	doi = {10.56801/10.56801/v8.i.452},
	language = {en},
	urldate = {2024-06-04},
	journal = {Journal of Modern Applied Statistical Methods},
	author = {Sawilowsky, Shlomo S.},
	month = nov,
	year = {2009},
	pages = {597--599},
}

@article{scheme_training_2013,
	title = {Training {Strategies} for {Mitigating} the {Effect} of {Proportional} {Control} on {Classification} in {Pattern} {Recognition}–{Based} {Myoelectric} {Control}},
	volume = {25},
	issn = {1040-8800},
	url = {https://journals.lww.com/00008526-201304000-00004},
	doi = {10.1097/JPO.0b013e318289950b},
	language = {en},
	number = {2},
	urldate = {2024-06-20},
	journal = {JPO Journal of Prosthetics and Orthotics},
	author = {Scheme, Erik and Englehart, Kevin},
	month = apr,
	year = {2013},
	pages = {76--83},
}

@article{robertson_effects_2019,
	title = {Effects of {Confidence}-{Based} {Rejection} on {Usability} and {Error} in {Pattern} {Recognition}-{Based} {Myoelectric} {Control}},
	volume = {23},
	copyright = {https://ieeexplore.ieee.org/Xplorehelp/downloads/license-information/IEEE.html},
	issn = {2168-2194, 2168-2208},
	url = {https://ieeexplore.ieee.org/document/8516279/},
	doi = {10.1109/JBHI.2018.2878907},
	number = {5},
	urldate = {2024-06-20},
	journal = {IEEE Journal of Biomedical and Health Informatics},
	author = {Robertson, Jason W. and Englehart, Kevin B. and Scheme, Erik J.},
	month = sep,
	year = {2019},
	pages = {2002--2008},
}

@article{campbell_current_2020,
	title = {Current {Trends} and {Confounding} {Factors} in {Myoelectric} {Control}: {Limb} {Position} and {Contraction} {Intensity}},
	volume = {20},
	copyright = {http://creativecommons.org/licenses/by/3.0/},
	issn = {1424-8220},
	shorttitle = {Current {Trends} and {Confounding} {Factors} in {Myoelectric} {Control}},
	url = {https://www.mdpi.com/1424-8220/20/6/1613},
	doi = {10.3390/s20061613},
	language = {en},
	number = {6},
	urldate = {2024-06-21},
	journal = {Sensors},
	author = {Campbell, Evan and Phinyomark, Angkoon and Scheme, Erik},
	month = jan,
	year = {2020},
	pages = {1613},
}

@article{woods_age-related_2015,
	title = {Age-related slowing of response selection and production in a visual choice reaction time task},
	volume = {9},
	issn = {1662-5161},
	url = {https://www.frontiersin.org/journals/human-neuroscience/articles/10.3389/fnhum.2015.00193/full},
	doi = {10.3389/fnhum.2015.00193},
	language = {English},
	urldate = {2024-06-27},
	journal = {Frontiers in Human Neuroscience},
	author = {Woods, David L. and Wyma, John M. and Yund, E. William and Herron, Timothy J. and Reed, Bruce},
	month = apr,
	year = {2015},
}

@article{soukoreff_towards_2004,
	series = {Fitts' law 50 years later: applications and contributions from human-computer interaction},
	title = {Towards a standard for pointing device evaluation, perspectives on 27 years of {Fitts}’ law research in {HCI}},
	volume = {61},
	issn = {1071-5819},
	url = {https://www.sciencedirect.com/science/article/pii/S1071581904001016},
	doi = {10.1016/j.ijhcs.2004.09.001},
	number = {6},
	urldate = {2024-07-01},
	journal = {International Journal of Human-Computer Studies},
	author = {Soukoreff, R. William and MacKenzie, I. Scott},
	month = dec,
	year = {2004},
	pages = {751--789},
}

@article{scheme_motion_2014,
	title = {Motion {Normalized} {Proportional} {Control} for {Improved} {Pattern} {Recognition}-{Based} {Myoelectric} {Control}},
	volume = {22},
	issn = {1558-0210},
	url = {https://ieeexplore.ieee.org/abstract/document/6473893/},
	doi = {10.1109/TNSRE.2013.2247421},
	number = {1},
	urldate = {2024-07-01},
	journal = {IEEE Transactions on Neural Systems and Rehabilitation Engineering},
	author = {Scheme, Erik and Lock, Blair and Hargrove, Levi and Hill, Wendy and Kuruganti, Usha and Englehart, Kevin},
	month = jan,
	year = {2014},
	pages = {149--157},
}

@article{sheehe_latin_1961,
	title = {Latin {Squares} to {Balance} {Immediate} {Residual}, and {Other} {Order}, {Effects}},
	volume = {17},
	issn = {0006-341X},
	url = {https://www.jstor.org/stable/2527834},
	doi = {10.2307/2527834},
	number = {3},
	urldate = {2024-07-01},
	journal = {Biometrics},
	author = {Sheehe, Paul R. and Bross, Irwin D. J.},
	year = {1961},
	pages = {405--414},
}

@article{smith_real-time_2014,
	title = {Real-time simultaneous and proportional myoelectric control using intramuscular {EMG}},
	volume = {11},
	copyright = {http://iopscience.iop.org/info/page/text-and-data-mining},
	issn = {1741-2560, 1741-2552},
	url = {https://iopscience.iop.org/article/10.1088/1741-2560/11/6/066013},
	doi = {10.1088/1741-2560/11/6/066013},
	number = {6},
	urldate = {2024-07-18},
	journal = {Journal of Neural Engineering},
	author = {Smith, Lauren H and Kuiken, Todd A and Hargrove, Levi J},
	month = dec,
	year = {2014},
	pages = {066013},
}

@article{xu_cross-user_2024,
	title = {Cross-{User} {Electromyography} {Pattern} {Recognition} {Based} on a {Novel} {Spatial}-{Temporal} {Graph} {Convolutional} {Network}},
	volume = {32},
	issn = {1558-0210},
	url = {https://ieeexplore.ieee.org/abstract/document/10354428},
	doi = {10.1109/TNSRE.2023.3342050},
	urldate = {2024-08-08},
	journal = {IEEE Transactions on Neural Systems and Rehabilitation Engineering},
	author = {Xu, Mengjuan and Chen, Xiang and Ruan, Yuwen and Zhang, Xu},
	year = {2024},
	pages = {72--82},
}

@article{wang_unravelling_2024,
	title = {Unravelling {Influence} {Factors} in {Pattern} {Recognition} {Myoelectric} {Control} {Systems}: {The} {Impact} of {Limb} {Positions} and {Electrode} {Shifts}},
	volume = {24},
	copyright = {http://creativecommons.org/licenses/by/3.0/},
	issn = {1424-8220},
	shorttitle = {Unravelling {Influence} {Factors} in {Pattern} {Recognition} {Myoelectric} {Control} {Systems}},
	url = {https://www.mdpi.com/1424-8220/24/15/4840},
	doi = {10.3390/s24154840},
	language = {en},
	number = {15},
	urldate = {2024-08-22},
	journal = {Sensors},
	author = {Wang, Bingbin and Li, Jinglin and Hargrove, Levi and Kamavuako, Ernest Nlandu},
	month = jan,
	year = {2024},
	pages = {4840},
}

@article{stuttaford_reducing_2024,
	title = {Reducing {Motor} {Variability} {Enhances} {Myoelectric} {Control} {Robustness} {Across} {Untrained} {Limb} {Positions}},
	volume = {32},
	issn = {1558-0210},
	url = {https://ieeexplore.ieee.org/abstract/document/10360873},
	doi = {10.1109/TNSRE.2023.3343621},
	urldate = {2024-08-22},
	journal = {IEEE Transactions on Neural Systems and Rehabilitation Engineering},
	author = {Stuttaford, Simon A. and Dyson, Matthew and Nazarpour, Kianoush and Dupan, Sigrid S. G.},
	year = {2024},
	pages = {23--32},
}

@article{zhang_multi-source_2023,
	title = {Multi-source domain generalization and adaptation toward cross-subject myoelectric pattern recognition},
	volume = {20},
	issn = {1741-2560, 1741-2552},
	url = {https://iopscience.iop.org/article/10.1088/1741-2552/acb7a0},
	doi = {10.1088/1741-2552/acb7a0},
	number = {1},
	urldate = {2024-08-22},
	journal = {Journal of Neural Engineering},
	author = {Zhang, Xuan and Wu, Le and Zhang, Xu and Chen, Xiang and Li, Chang and Chen, Xun},
	month = feb,
	year = {2023},
	pages = {016050},
}

@inproceedings{jabbari_emg-based_2020,
	title = {{EMG}-{Based} {Hand} {Gesture} {Classification} with {Long} {Short}-{Term} {Memory} {Deep} {Recurrent} {Neural} {Networks}},
	url = {https://ieeexplore.ieee.org/abstract/document/9175279},
	doi = {10.1109/EMBC44109.2020.9175279},
	urldate = {2024-08-26},
	booktitle = {2020 42nd {Annual} {International} {Conference} of the {IEEE} {Engineering} in {Medicine} \& {Biology} {Society} ({EMBC})},
	author = {Jabbari, Milad and Khushaba, Rami N. and Nazarpour, Kianoush},
	month = jul,
	year = {2020},
	note = {ISSN: 2694-0604},
	pages = {3302--3305},
}

@article{zhai_self-recalibrating_2017,
	title = {Self-{Recalibrating} {Surface} {EMG} {Pattern} {Recognition} for {Neuroprosthesis} {Control} {Based} on {Convolutional} {Neural} {Network}},
	volume = {11},
	issn = {1662-453X},
	url = {https://www.frontiersin.org/journals/neuroscience/articles/10.3389/fnins.2017.00379/full},
	doi = {10.3389/fnins.2017.00379},
	language = {English},
	urldate = {2024-08-26},
	journal = {Frontiers in Neuroscience},
	author = {Zhai, Xiaolong and Jelfs, Beth and Chan, Rosa H. M. and Tin, Chung},
	month = jul,
	year = {2017},
}

@article{li_gesture_2021,
	title = {Gesture {Recognition} {Using} {Surface} {Electromyography} and {Deep} {Learning} for {Prostheses} {Hand}: {State}-of-the-{Art}, {Challenges}, and {Future}},
	volume = {15},
	issn = {1662-453X},
	shorttitle = {Gesture {Recognition} {Using} {Surface} {Electromyography} and {Deep} {Learning} for {Prostheses} {Hand}},
	url = {https://www.frontiersin.org/journals/neuroscience/articles/10.3389/fnins.2021.621885/full},
	doi = {10.3389/fnins.2021.621885},
	language = {English},
	urldate = {2024-08-26},
	journal = {Frontiers in Neuroscience},
	author = {Li, Wei and Shi, Ping and Yu, Hongliu},
	month = apr,
	year = {2021},
}

@inproceedings{ortiz-catalan_offline_2015,
	title = {Offline accuracy: {A} potentially misleading metric in myoelectric pattern recognition for prosthetic control},
	shorttitle = {Offline accuracy},
	url = {https://ieeexplore.ieee.org/abstract/document/7318567},
	doi = {10.1109/EMBC.2015.7318567},
	urldate = {2024-08-26},
	booktitle = {2015 37th {Annual} {International} {Conference} of the {IEEE} {Engineering} in {Medicine} and {Biology} {Society} ({EMBC})},
	author = {Ortiz-Catalan, Max and Rouhani, Faezeh and Brånemark, Rickard and Håkansson, Bo},
	month = aug,
	year = {2015},
	note = {ISSN: 1558-4615},
	pages = {1140--1143},
}

@article{raghu_enabling_2024,
	title = {Enabling Myoelectirc Control Training Using Continuous Data Through Self-Supervised Representation Learning},
	copyright = {Copyright (c) 2024 MEC Symposium Conference},
	url = {https://conferences.lib.unb.ca/index.php/mec/article/view/2503},
	doi = {10.57922/mec.2503},
	urldate = {2024-09-06},
	journal = {Myoelectric Controls Symposium},
	author = {Raghu, Shriram Tallam Puranam and MacIsaac, Dawn and Scheme, Erik},
	month = aug,
	year = {2024},
}

@article {eddy_bigdata,
	author = {Eddy, Ethan and Campbell, Evan and Bateman, Scott and Scheme, Erik},
	title = {Big Data in Myoelectric Control: Large Multi-User Models Enable Robust Zero-Shot EMG-based Discrete Gesture Recognition},
	year = {2024},
	doi = {10.1101/2024.07.11.603119},
	URL = {https://www.biorxiv.org/content/early/2024/07/29/2024.07.11.603119},
	journal = {bioRxiv}
}

@INPROCEEDINGS{campbell_ciil,
  author={Eddy, Ethan and Campbell, Evan and Bateman, Scott and Scheme, Erik},
  booktitle={2023 IEEE International Conference on Systems, Man, and Cybernetics (SMC)}, 
  title={Leveraging Task-Specific Context to Improve Unsupervised Adaptation for Myoelectric Control}, 
  year={2023},
  volume={},
  number={},
  pages={4661-4666},
  doi={10.1109/SMC53992.2023.10394393}}

@article{williams_multifaceted_2024,
	title = {A multifaceted suite of metrics for comparative myoelectric prosthesis controller research},
	volume = {19},
	issn = {1932-6203},
	url = {https://journals.plos.org/plosone/article?id=10.1371/journal.pone.0291279},
	doi = {10.1371/journal.pone.0291279},
	language = {en},
	number = {5},
	urldate = {2024-09-16},
	journal = {PLOS ONE},
	author = {Williams, Heather E. and Shehata, Ahmed W. and Cheng, Kodi Y. and Hebert, Jacqueline S. and Pilarski, Patrick M.},
	month = may,
	year = {2024},
	pages = {e0291279},
}

@article{mukhopadhyay_experimental_2020,
	title = {An experimental study on upper limb position invariant {EMG} signal classification based on deep neural network},
	volume = {55},
	issn = {1746-8094},
	url = {https://www.sciencedirect.com/science/article/pii/S1746809419302502},
	doi = {10.1016/j.bspc.2019.101669},
	urldate = {2024-09-16},
	journal = {Biomedical Signal Processing and Control},
	author = {Mukhopadhyay, Anand Kumar and Samui, Suman},
	month = jan,
	year = {2020},
	pages = {101669},
}

@article{eddy_libemg:_2023,
	title = {{LibEMG}: {An} {Open} {Source} {Library} to {Facilitate} the {Exploration} of {Myoelectric} {Control}},
	volume = {11},
	issn = {2169-3536},
	shorttitle = {{LibEMG}},
	doi = {10.1109/ACCESS.2023.3304544},
	urldate = {2024-09-14},
	journal = {IEEE Access},
	author = {Eddy, Ethan and Campbell, Evan and Phinyomark, Angkoon and Bateman, Scott and Scheme, Erik},
	year = {2023},
	pages = {87380--87397},
}

@misc{raghu_self-supervised_2024,
	title = {Self-{Supervised} {Learning} via {VICReg} {Enables} {Training} of {EMG} {Pattern} {Recognition} {Using} {Continuous} {Data} with {Unclear} {Labels}},
	url = {http://arxiv.org/abs/2409.11632},
	doi = {10.48550/arXiv.2409.11632},
	urldate = {2024-09-19},
	publisher = {arXiv},
	author = {Raghu, Shriram Tallam Puranam and MacIsaac, Dawn T. and Scheme, Erik J.},
	month = sep,
	year = {2024},
	note = {arXiv:2409.11632 [eess]},
}

@inproceedings{NEURIPS2019_a2b15837,
 author = {Hendrycks, Dan and Mazeika, Mantas and Kadavath, Saurav and Song, Dawn},
 booktitle = {Advances in Neural Information Processing Systems},
 editor = {H. Wallach and H. Larochelle and A. Beygelzimer and F. d\textquotesingle Alch\'{e}-Buc and E. Fox and R. Garnett},
 pages = {},
 publisher = {Curran Associates, Inc.},
 title = {Using Self-Supervised Learning Can Improve Model Robustness and Uncertainty},
 url = {https://proceedings.neurips.cc/paper_files/paper/2019/file/a2b15837edac15df90721968986f7f8e-Paper.pdf},
 volume = {32},
 year = {2019}
}
\end{document}